\documentclass[universe,article,accept,pdftex,moreauthors]{Definitions/mdpi} 

\firstpage{1} 
\pubvolume{1}
\issuenum{1}
\articlenumber{0}
\pubyear{2026}
\copyrightyear{2026}
\externaleditor{Firstname \\Lastname}
\datereceived{25 March 2026} 
\daterevised{25 June 2026} 
\dateaccepted{6 July 2026} 
\datepublished{ } 
\hreflink{https://doi.org/} 

\usepackage{amsfonts,amssymb,amsmath,graphicx,multirow,dsfont}

\makeatletter
\def\maketag@@@#1{\hbox{\m@th\normalfont\normalsize#1}}
\makeatother

\graphicspath{{figures/}}
\def\dac{\displaystyle\frac}

\def\[{\left[}
\def\]{\right]}
\def\({\left(}
\def\){\right)}

\Title{On the Stability of  Dark Energy Scalar Field Reconstruction from SNe Ia Data}

\Author{{{Arpine Piloyan} $^{1,2,\dagger}$\orcidA{} and Sergey Pavluchenko $^{2, 3,}$*$^{,\dagger}$\orcidB{}} }

\AuthorNames{Arpine Piloyan and Sergey Pavluchenko}

\address{%
$^{1}$ \quad A. I. Alikhanian National Science Laboratory (Yerevan Physics Institute), Alikhanyan Broth. 2, \mbox{Yerevan {0036}, Armenia;} {arpine.piloyan@gmail.com}
\\
$^{2}$ \quad Programa de P\'os-Gradua\c{c}\~ao em F\'isica, Universidade Federal do Maranh\~ao (UFMA), \mbox{S\~ao Lu\'is, MA 65085-580, Brazil}\\
$^{3}$ \quad {Department of Bioengineering}
, University of Illinois Urbana-Champaign, Urbana, IL {61801}, USA}
 
\corres{Correspondence: sergey.pavluchenko@gmail.com}

\firstnote{These authors contributed equally to this work.}

\abstract{The current paper addresses the possibility of  Dark Energy scalar field potential reconstruction from  SNe Ia data and the problems arising during the process. We describe the method and test its limits and features with use of synthetic data, as well as discuss several issues connected to error propagation. We conclude that the chosen smoothing method---binning of the data---introduces immense uncertainty amplification which limits the practical application of this method, leaving us with other alternatives. 
We also address the ``instability of the reconstruction'', an important effect when a false scalar field---real or phantom---could be reconstructed if the $\Omega_m$ and $H_0$ parameters are wrongly estimated; a similar effect could be expected in other Dark Energy models as well.}


\begin{document}

\section{Introduction}

The discovery of the accelerated expansion of the Universe in 1998~\cite{riess,perl} is an important milestone of  modern cosmology. This behavior is quite different from the Friedmann model,
which is expected from the Standard Cosmology description, developed in the 1960s--1970s. The Friedmann model description assumes only ordinary matter sources, and leads to power-law behavior, so that to fit the observed
evolution, we need to add ``unusual'' matter, or to modify General Relativity (GR). The class of matter sources which could lead to the accelerated expansion is called Dark Energy (DE), as an energy source of unknown
origin, and corresponding  to the naming of Dark Matter (DM).

Of course, this behavior requires an explanation---and preferably a natural one---and, as a
consequence, it has led to a rapid increase in suggested possible causes. The simplest of the DE models is just the cosmological constant ($\Lambda$-term), but this approach has its own difficulties~\cite{carrol, pad, ss00}.
Other approaches
include phantom cosmology~\cite{rev_add1, rev_add1.1}, tachyonic matter~\cite{rev_add2, rev_add2.2}, braneworld scenarios~\cite{rev_add3, rev_add3.3},
scalar-tensor theories~\cite{rev_add4, rev_add4_1, rev_add4.4}, $f(R)$-gravity~\mbox{\cite{rev_add5, rev_add5.1, rev_add5.2}}, holographic gravity~\cite{rev_add6, rev_add7, rev_add7.1}, Chaplygin gas~\cite{rev_add8, rev_add9, rev_add10, rev_add11, rev_add2.2},
models with extra (i.e., more than four) dimensions~\cite{rev_add12}, neutrinos of varying mass~\cite{rev_add13, rev_add14, rev_add14.1}, multi-coupled DE scalar fields~\cite{addd1, addd2} and many others.
In addition, there are also models with fluid-based approaches (e.g.,~\cite{rev_add14.5}) or
Cardassian cosmology~\cite{rev_add15, rev_add16, rev_add17}. One may note that many of the theoretical models could be united under the approach of  Horndeski's theory~\cite{horndeski},
which builds the most general ghost-free second-order Lagrangian. It includes some of the above-mentioned theories, but not all---for instance, it includes scalar fields and $f(R)$, but not $f(\mathcal{G})$, the
generalization of $f(R)$ for the Gauss--Bonnet term. In addition to the references above, we would like to recommend~\cite{WP} as a comprehensive review of the current status of DE models---mostly observational, though.

In our work, we are using a scalar field to describe the accelerated expansion. In a sense, it is similar to the inflationary approach, but with a number of differences, like the energy scale (for inflation,
we require Planck energy, while for current expansion, it is much lower) and other features (for instance, inflation should end and there should be a mechanism for inflaton decay, while for the scalar field responsible
for current acceleration, there are no such requirements); one can even describe both inflation and DE with the same potentials~\cite{prd03}. 

Without doubt, this is not the first attempt to reconstruct the DE scalar field---see, e.g.,~\mbox{\cite{starobinsky, ellis, turner, saini}} as examples of early reconstruction
attempts. However, the cited papers used one or another parameterization---of either the equation of state of the DE $\omega(z)$ or the Hubble parameter $H(z)$ or some other quantity---with a small number of free parameters. We can
also note that similar reconstructions for some general parameterized DE can be found (see, e.g.,~\cite{add1, add2, add3, add4, add5, add6, add7, add8, add9, add10}).

The above-cited works, as well as the current investigation,  use supernovae Ia (SNe Ia) data. 
This is a good choice---it is agreed that the SNe Ia are ``standard candles'' (with a certain precision) which
means that they have the same (known) luminosity and by measuring the flux one can obtain the distance, known as the photometric distance $D_L$, to them.
On the other hand, spectroscopic observations provide us with the redshift $z$ for each particular SNe. Having both the photometric distance $D_L$ and the redshift $z$, one can build the so-called comoving distance
$D_M = D_L / (1+z)$ and reconstruct $H(z)$. The reconstructed $H(z)$ is the starting point for all DE reconstructions---for instance, in our scalar field case, the potential and the kinetic terms are reconstructed
from $H(z)$ and its derivative $H'(z) \equiv dH(z)/dz$; the other approaches are quite similar. If $\omega$CDM assumes parameterization of the equation of state of the Dark Energy in the form \mbox{$\omega=\omega_{0}+\omega_{1}(1+z)+\ldots$}, it could be connected to our considered scalar field case quite easily: in the simplest case, $\omega = \omega_0 \ne -1$, the Dark Energy could be described in terms of scaling~\cite{sc1, sc2} or tracker~\cite{tr} solutions---exponential, inverse power-law and several others~\cite{tr}.
With more terms in the $\omega$ decomposition, the solutions for the scalar field become less likely to be found, especially in the closed form, which makes the connection of $\omega$CDM with scalar fields less obvious.
Clearly, the potential, resulting from $\omega$CDM, is a simple exponential or power-law potential that depends only on one or two parameters. In this case, the reconstruction will be much less noisy than in our
parameter-free case.
Similarly, the cosmography approach (see~\cite{cosmo1} for review;~\cite{cosmo2, cosmo3, cosmo4, cosmo5} for the effects on GR level;~\cite{cosmo6} for extended gravity;
\cite{cosmo7} for $f(R)$ gravity; and \cite{cosmo8, cosmo9} for $f(T)$ gravity)
uses higher-order derivatives of the Hubble parameter: $H''(z)$, $H'''(z)$ and others. One can see that both of them are quite similar in  approach but parameterize different variables.

We mentioned above that the previous DE scalar field reconstructions used some sort of parameterization while in our work we keep all parameters free. As we discussed in~\cite{pre_proc}, the models with and
without parameterization pursue different goals and, as a consequence, have different results. With the use of parameterization, one usually has much lower errors---indeed, with parameterized $H(z)$ or $\omega(z)$, it is
much easier to reconstruct  all the necessary quantities (sometimes even in a semi-analytical way), but particular parameterization immediately biases the scalar field potential shape (for our case). By contrast,
without use of any parameterization, the errors of the reconstruction are several-fold higher, but the results do not have any bias (see~\cite{pre_proc} \mbox{for details).}

The current manuscript extends the analysis performed in~\cite{pre_proc}. 
There, we reported the technical details of the reconstruction process while we now undertake
a full-scale analysis of the entire process. The manuscript has three goals---to demonstrate that the proposed scheme works correctly, to demonstrate that the error propagation leads to unreasonably high uncertainties due to employment of the binning scheme, and to investigate the effect of the usage of wrong values of $H_0$ and $\Omega_m$ on the reconstructed potential (``(in)stability of the reconstruction'').
The structure of the manuscript is as follows:
	In Section~\ref{potslope}, we present the basic equations and show how the $V(\phi)$ potential could be reconstructed using Friedman equations for the Friedmann--Robertson--Walker (FRW) metrics; in the same section, we also present the reconstruction techniques and methods.
After that, we test our reconstruction method on  synthetic data---$\Lambda$CDM in Section~\ref{syndata} and the exponential potential in Section~\ref{sc.f.}. Error propagation
 and its effects on the reconstruction are analyzed in Section~\ref{s.err} while analysis of the reconstruction stability is performed in Section~\ref{s.stab}. Finally, we discuss the obtained results in Section~\ref{s.discuss} and draw conclusions in Section~\ref{s.con}.

\section{{Potential Reconstruction}}
\label{potslope}
	First of all, we must demonstrate the possibility of scalar field reconstruction without any assumptions. Thereby, let us show that there always exists a single field potential $V(\phi)$ that reproduces any observed Hubble parameter as a function of the redshift, $H(z)$. We also need to demonstrate that the scalar field $\phi(z)$ could also be reconstructed from $H(z)$.
The only general assumptions are that the Universe contains matter with a known generic constant
equation of state $w_{m}$ and a single canonical minimally coupled
scalar field.
The well-known two independent Friedmann {equations are} (we put $8\pi G\equiv 1$)
\begin{align}
3H^{2} & =\rho_{m}+\frac{\dot{\phi}^{2}}{2}+V(\phi)\label{eq:v}\\
2\dot{H} & =-\dot{\phi}^{2}-\rho_{m}(1+w_{m})\label{eq:hd}
\end{align}

Adopting the redshift $z$ as an affine parameter, we can write from (\ref{eq:hd})
\begin{equation}
H(z)^{2}(1+z)^{2}\left(\frac{d\phi}{dz}\right)^{2}=2(1+z)H(z)\frac{dH}{dz}-\rho_{m}(z)(1+w_{m})\label{eq:pdd}
\end{equation}
where $\rho_{m}(z)$ is the solution of
\begin{equation}
(1+z)\frac{d\rho_{m}}{dz}=3\rho_{m}(z)(1+w_{m}).\label{eq:pdd1}
\end{equation}

Obtaining $\rho_m$ from (\ref{eq:pdd1}) and substituting it to (\ref{eq:pdd}), after integration, we can derive $\phi(z)$.
 Then, by combining (\ref{eq:pdd})
and (\ref{eq:v}), with use of (\ref{eq:pdd1}), one obtains
\begin{equation}
V(z)=3H(z)^{2}-(1+z)H(z)\frac{dH(z)}{dz}+\frac{\rho_{m}(z)}{2}(w_{m}-1).
\end{equation}

Finally, by jointly using $V(z)$ and $\phi(z)$, it is possible to reconstruct $V(\phi)$
for any observed $H(z)$. There is no guarantee, however, that the formal
solution obtained this way is stable or unique or free of singularities.

Under the assumption that $w_m = 0$ (the pressureless matter), the solution of (\ref{eq:pdd1}) could be written in the form $\Omega_m = \Omega_m^0 (1+z)^3$, where $\Omega_m^0$ is the  current value for the fraction of  ordinary matter in the total energy budget. After we also introduce $H_0$, the current value for the Hubble parameter, the equations can be rewritten in the following way:
\vspace{-1pt}
\begin{equation}
\tilde V \equiv \dac{8\pi G}{3 H_0^2} V(z) = \dac{H(z)^2}{H_0^2} - \dac{H(z) H'(z) (1+z)}{3 H_0^2} - \dac{\Omega_m^0 (1+z)^3}{2}, \\ \\
\label{Vzn}
\end{equation}
\begin{equation}
\tilde{\( \dac{d\phi}{dz} \)^2} \equiv \dac{8\pi G}{3 H_0^2} \( \dac{d\phi}{dz} \)^2 = \dac{2 H'(z)}{3 H(z) H_0^2 (1+z)} - \dac{\Omega_m^0 (1+z)}{H(z)^2}, \\ \\
\label{dfdzn}
\end{equation}

Our next step is to show how $H(z)$ and its derivatives can be reconstructed from data, namely, from SNe type Ia for which as an input, we have the distance modulus $\upmu_i$ with its error $\delta\upmu_i$ for each $i$th supernova at redshift $z_i$ (maybe with error $\delta z_i$, depending on the data).

 For smoothing of the individual values of the distance modulus, we use \mbox{two binning} methods---binning with equal $\Delta z$ and with equal $\Delta N$ (number of supernovae), as well as their combination, if necessary. Later, we will discuss the
 {\emph{pros}} 
 and {\emph{cons}} of these methods.
Because we assume the Gaussian nature of the SNe errors, in each bin, the average values for the distance modulus
 are $\overline\upmu_j = (\sum_i \upmu_i)/N_j$ with an error $\sigma_{\overline\upmu_j} = \sqrt{(\sum_i \sigma_i^2)/N_j}$, where $N_j$ is the number of supernovae in the $j$th bin. We also define $\delta z$---the error in $z$
as the half-width of the bin (in  case we have $\delta z_i$ for each individual supernova, the definition for the error in binned $z$ becomes more complicated). We keep this notation for the general case as for equal $z$ binning, it gives the same value for $\delta z$, but for alternative binning, it will \mbox{be different.}

Firstly, we transform the distance modulus and its errors into the comoving distance $D_M$:
\begin{equation}
\begin{array}{l}
D_M = \dac{10^{\(\dac{\upmu}{5} + 1\)}}{1+z}, ~~\delta D_M = \delta \( \dac{10^{\(\dac{\upmu}{5} + 1\)}}{1+z} \) = \cdots = D_M \( \dac{\ln 10\delta\upmu}{5} + \dac{\delta z}{(1+z)}  \).
\end{array} \label{dL}
\end{equation}

\noindent {We process} further to $H(z)$, which is defined as $H(z) = \( d D_M/dz\)^{-1}$.
We use a one-step differentiation scheme for simplicity, then the value and the error take the form

\begin{equation}
\begin{array}{l}
H(z) = \dac{1}{D_M'},~~ \delta H(z) = \delta \( \dac{1}{D_M'}  \)  = \dac{\delta D_M'}{(D_M')^2}, ~~ D_M' = \dac{D_{M, 2} - D_{M, 1}}{z_2 - z_1}, \\ \\
\delta (D_M') = \delta \( \dac{D_{M, 2} - D_{M, 1}}{z_2 - z_1}  \) = \cdots = \dac{\delta D_{M, 2} + \delta D_{M, 1}}{z_2 - z_1} + |D_M'| \dac{\delta z_2 + \delta z_1}{z_2 - z_1}.
\end{array} \label{Hz}
\end{equation}

\noindent  {Let us also} calculate the
derivative of $H(z)$ with respect to $z$:
\begin{equation}
\begin{array}{l}
H' = \dac{H_2 - H_1}{z_2 - z_1},~~\delta H' = \cdots = \dac{\delta H_2 + \delta H_1}{z_2 - z_1} + |H'| \dac{\delta z_2 + \delta z_1}{z_2 - z_1}.
\end{array} \label{dH}
\end{equation}

Now, with both $H(z)$ (see Equation (\ref{Hz})) and $H'$ (see Equation (\ref{dH})) calculated, we can recover the potential $V(z)$ and the kinetic part $(d\phi/dz)^2$. Errors can be calculated from Equations (\ref{Vzn}) and (\ref{dfdzn}), in the following way:
\begin{equation}
\delta \tilde V = \dac{2H\delta H}{H_0^2} + \dac{(1+z)H'\delta H + H(1+z)\delta H' + HH'\delta z}{3H_0^2} + \dac{3(1+z)^2 \Omega_m^0 \delta z}{2}; \label{Vz}
\end{equation}
\small
\begin{equation}
\delta \( \tilde{\( \dac{d\phi}{dz} \)^2} \) = \dac{2}{3H_0^2} \[ \dac{\delta H'}{H(1+z)} + \dac{H'\delta H}{H^2(1+z)} + \dac{H' \delta z}{H(1+z)^2} \] + \Omega_m^0 \( \dac{\delta z}{H^2} + \dac{2(1+z)\delta H}{H^3}  \). \label{dfdz2}
\end{equation}

\vspace{6pt}

 \normalsize The Equation (\ref{dfdzn}) (if positive!) could be integrated to get $\phi(z)$ (with an additive constant!).
 We use the rectangle method for integration; thus, the error propagation for the remaining steps
is as follows:
\vspace{-1pt}
\begin{equation}
\begin{array}{l}
\dac{d\phi}{dz} = \sqrt{(\dac{d\phi}{dz})^2} \Rightarrow \delta \( \dac{d\phi}{dz} \) = \delta \( \( \dac{d\phi}{dz} \)^2 \) / \( 2 \dac{d\phi}{dz}\); \\  \\
\phi = \int \( \dac{d\phi}{dz} \) dz \cong \( \dac{d\phi}{dz} \)|_{z_{central}} \Delta z \Rightarrow \delta\phi = \Delta z \times \delta \( \dac{d\phi}{dz}\).
\end{array} \label{rest}
\end{equation}

Finally, using $V(z)$ and $\phi(z)$ together, we can recover $V(\phi)$.

At this point, it is appropriate to mention an ambiguity within the potential reconstruction, rooted in two steps. The first of the sources originates from (\ref{dfdzn})---we reconstruct the kinetic term which is quadratic, which means that the field itself could either rise or decrease with $z$ (so that in the first of  (\ref{rest}), there should be $\pm$ in front of the square root). The second source derives from the fact that when we integrate the kinetic term to get the field itself, the reconstruction is valid up to a null-point (integration constant). So, we actually reconstruct the shape of the potential which is subject to shifts along the $\phi$-axis as well as mirroring to account for rising/falling potential. Due to the reconstruction technique, it is possible to reconstruct the potential only up to this ambiguity, and to get the actual potential, we have to rely on the additional assumptions or observational data.

In the sections to follow, we try the described scheme first on synthetic $\Lambda$CDM data and then on  synthetic data from the model with exponential potential.

\section{$\boldsymbol{\Lambda}$CDM Synthetic Data}
\label{syndata}
In this section, we test our scheme with synthetic $\Lambda$CDM data. We generate binned data for $\Lambda$CDM cosmology with $H_0 = 68$ km/s/Mpc and $\Omega_m = 0.25$ with bin size $\delta z = 0.025$ and $\delta\upmu = 0.5$,
process it and recover $V(z)$ and $(d\phi/dz)^2$. As it is $\Lambda$CDM, the data are generated using the standard cosmological relationships
\begin{equation}
\begin{array}{l}
H(z) = H_0 \sqrt{\Omega_m (1+z)^3 + \Omega_\Lambda} \\ \\
D_M(z) = \dac{c(1+z)}{H_0} \int_0^z \dac{dx}{\sqrt{\Omega_m (1+x)^3 + \Omega_\Lambda}}\\ \\
\upmu = 5 \log_{10} D_M + 25,
\end{array} \label{LCDM_gen}
\end{equation}
an appropriate rescaling and normalizations. Since our goal here is just to verify that $\Lambda$CDM recovers as a constant potential, we have not introduced additional variation of the central values with respect to $\Lambda$CDM predictions, as it will only contribute to the error budget. For the reference, the resulting data file used for the potential reconstruction for 
$\Lambda$CDM case is provided as ${\tt synth\_LCDM.dat}$ within the Supplementary Materials; the format is ($z$, $\delta z$, $\upmu$, $\delta \upmu$) with $\delta z = 0.025$ (half-bin size) and $\delta \upmu = 0.5$ (some accepted value; affects only error propagation).

\begin{figure}[H]
\includegraphics[width=11cm]{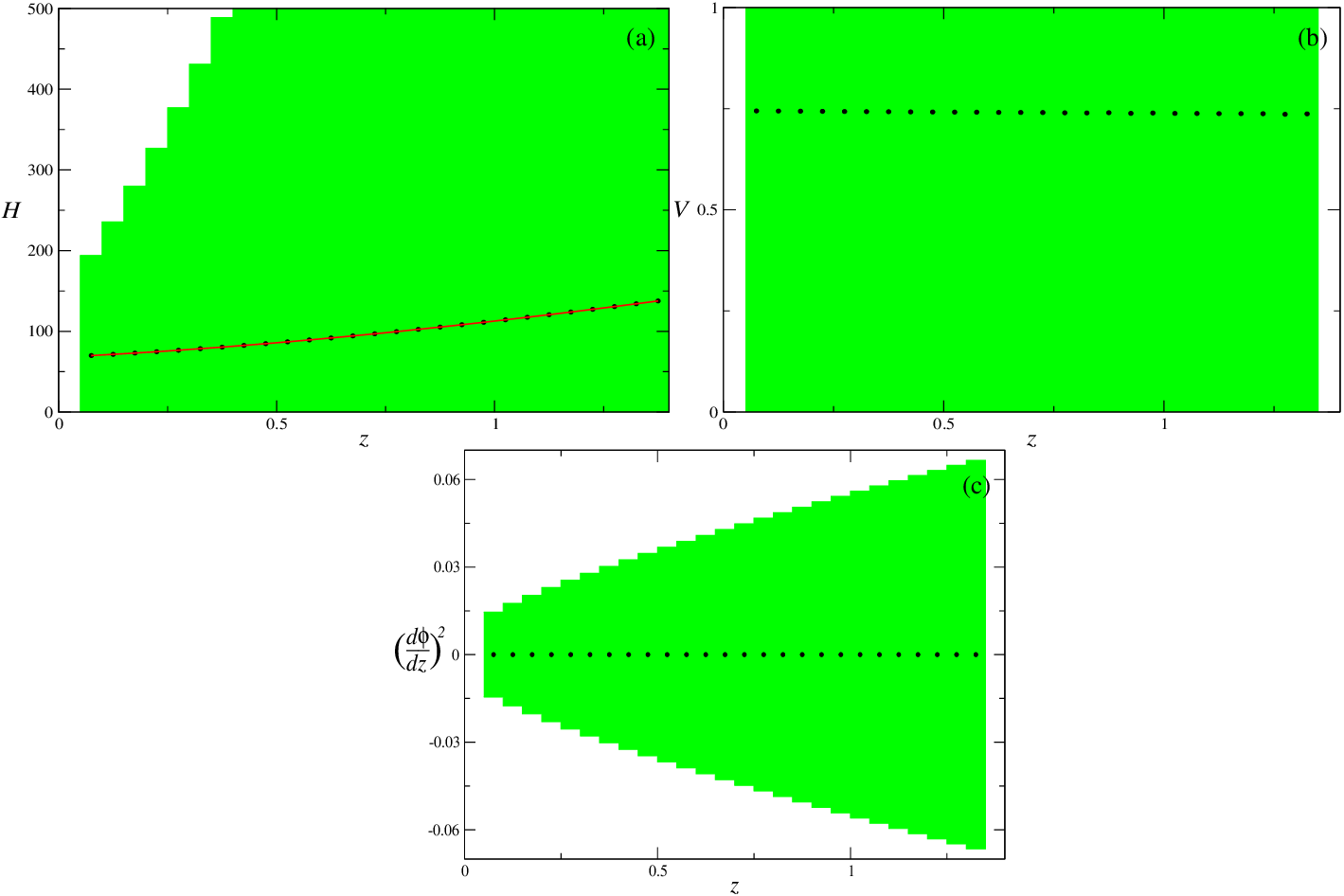}
\caption{{The results }of the potential recovery for the synthetic $\Lambda$CDM data. In all panels, the central values are depicted as black circles and the green area corresponds to $\pm 1\sigma$ from the central value in each bin.
In panel (\textbf{a}), we present the recovered $H(z)$ and add the theoretical curve as a red line. In  panel (\textbf{b}), we show the recovered $V(z)$ and in panel (\textbf{c}), the recovered $(d\phi/dz)^2$
(see the text for more details).}\label{fig_lcdm}
\end{figure}

The results are presented in Figure~\ref{fig_lcdm}. There, in panel (a), we present the results of the $H(z)$ recoverery---central values as black circles, green area as
the $\pm 1\sigma$ area and the red curve as the theoretical $H(z)$ curve calculated for $H_0 = 68$ km/s/Mpc and $\Omega_m = 0.25$. One can clearly see that the $H(z)$ recovery is perfect---the central values exactly coincide with the
theoretical curve. In panel (b), we present the recovered $V(z)$ with the same notations (except for the theoretical curve). As expected, for $\Lambda$CDM, the ``potential'' is just the constant $\Lambda$-term and with the
proper coefficient, it corresponds to the chosen $\Omega_\Lambda = 0.75$. Finally, in  panel (c), we present the results for the $(d\phi/dz)^2$
recovery. Again, as expected from the theory, for $\Lambda$CDM, it should be exactly zero and that is what we observe in Figure~\ref{fig_lcdm}c---the divergences are on the level of $10^{-12}$, which could be treated as
``exact zero''.

So, one can clearly see that for $\Lambda$CDM (and so for the constant potential), our scheme works perfectly; in the next section, we test it with real non-constant potential and see how the recovery works in that case.

\section{Exponential Potential Synthetic Data}
\label{sc.f.}

Now, it is time to test our scheme with synthetic data obtained from use of the exponential potential for the scalar field. Similarly to the previous section, we generate binned data with the
following parameters: $H_0 = 68$ km/s/Mpc, $\Omega_m = 0.25$ and with bin size $\delta z = 0.025$ and $\delta\upmu = 0.5$.
Since it is a Friedmann model with the scalar field, (\ref{LCDM_gen}) could not be used anymore and we have to solve the full system with the scalar field instead:
\begin{equation}
\begin{array}{l}
\dac{1}{4\pi G} \( \dot H + \dac{3}{2}H^2 \) + \( \dac{\dot\phi^2}{2} - V(\phi) \) = 0, \\ \\
\ddot\phi + 3H\dot\phi + \dac{dV}{d\phi} = 0,
\end{array} \label{exp_gen}
\end{equation}
and the initial conditions are found from the constrain equation
\begin{equation}
\begin{array}{l}
\dac{3H^2}{8\pi G}  = \( \dac{\dot\phi^2}{2} + V(\phi) + \dac{C_m}{a^3} \).
\end{array} \label{exp_constr}
\end{equation}

There, $H \equiv H(t)$ and $\phi \equiv \phi(t)$ are the Hubble parameter and scalar field as  functions of cosmic time $t$ and $C_m$ accounts for ordinary matter (normalized to meet the accepted $\Omega_m = 0.25$ at current time). Then, we start the numerical simulation at some time in the past and perform it until the contribution of the scalar field to the energy budget ($\Omega_{DE}$) meets the current accepted value ($\Omega_\Lambda = 1 - \Omega_m$) with $\Omega_{DE}$ given as
\begin{equation}
\begin{array}{l}
\Omega_{DE} = \dac{8\pi G}{3H^2} \( \dac{\dot\phi^2}{2} + V(\phi)  \).
\end{array} \label{omega_de}
\end{equation}

Then, we renormalize the scale factor to 1 and the Hubble parameter to the accepted value {$H_0 = 68~\text{km/s/Mpc}$} 
 and recover $H(z)$; for the reference, the resulting data file used for the potential reconstruction for 
the exponential potential case is provided as ${\tt synth\_exp.dat}$ within the Supplementary Materials; the exact form of
the potential is
\begin{equation}
\begin{array}{l}
V(\phi) = V_0 + A \exp(B\phi)
\end{array} \label{exp_pot}
\end{equation}
with $V_0 = (1 - \Omega_m) = 0.75$, $A = 1.67 \times 10^{-3}$ and $B=1832$. There, $A \sim M_p^{-4}$ (since we normalize everything to the critical density), which leads to the accepted numerical value after applying $8\pi G=1$ normalization; as for the value of $B$, in~\cite{sc1}, it was noted that for the exponential potential in the form of $V \propto \exp\(\lambda\kappa\phi\)$ with $\kappa^2 = 8\pi G$, in order to be consistent with the nucleosynthesis bound, one needs $\lambda^2 > 20$; to be conservative, we adopted an even higher value $\lambda \sim 60$ which leads to the accepted numerical value for $B$ after applying $8\pi G=1$ normalization; see also~\cite{exp.pot.2} for example of $\lambda = 20$.

\begin{figure}[H]
\includegraphics[width=12cm]{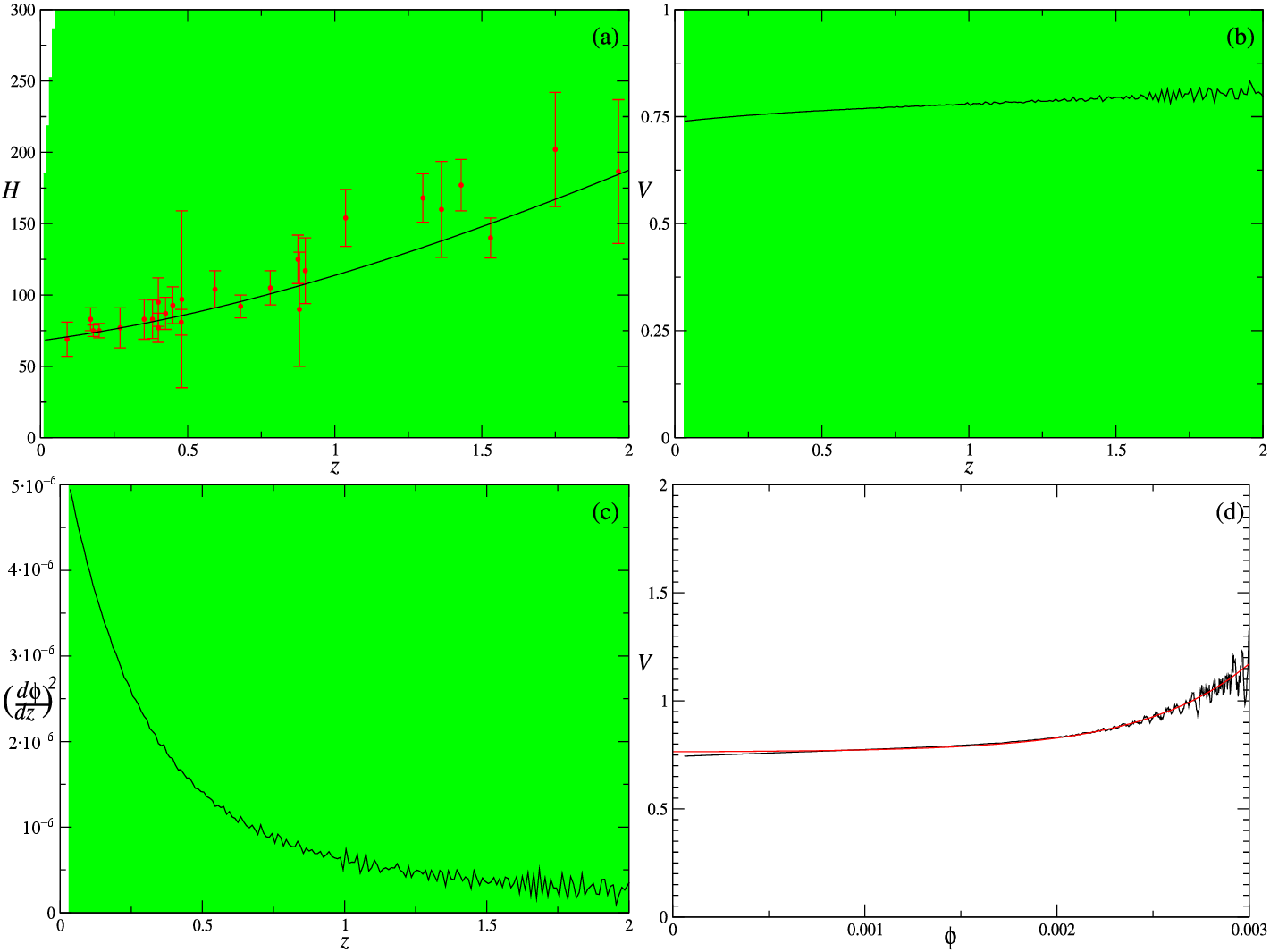}
\caption{{The results} of the potential reconstruction for the case of the exponential potential. Green area corresponds to the $\pm 1\sigma$ error budget; black curves correspond to the reconstructed
values.
In ~panel (\textbf{a}), we present the reconstructed $H(z)$ and non-SNe-based observational values for $H(z)$ as red points; in  panel (\textbf{b}), we present the reconstructed potential $V(z)$; in  panel (\textbf{c}), we
present the reconstructed $(d\phi/dz)^2$,  and finally, in panel (\textbf{d}),  we show the reconstructed potential $V(\phi)$ while the red curve indicates the theoretical potential
(see the text for more details).}\label{fig_exp}
\end{figure}

 Then, we process the data as described above and the intermediate and
final results are presented in Figure~\ref{fig_exp}. On all panels, the green area corresponds to the $\pm 1\sigma$ error budget; black lines show the reconstructed quantity. So, in  panel (a), we present
the reconstructed $H(z)$ curve and add available non-SNe-based measurements of $H(z)$ as red points. In  panel (b), we present the reconstructed $V(z)$ potential, while in panel (c), we present the reconstructed
kinetic term $(d\phi/dz)^2$. Please note that it is always positive---as we will see later, this is not always the case when dealing with realistic data.
After integrating the kinetic term, we obtain the scalar field $\phi (z)$ and by combining$V(z)$ with it, we finally obtain the ``true'' form of the potential $V(\phi)$ in panel (d).
There, we have omitted the error budget---from the previous panels, we can see that it is immense. Also, for comparison, we have provided the theoretical curve for the potential as a red line.

Bearing in mind the reconstruction ambiguity mentioned above, since it is mock data and its purpose is just a demonstration, we assumed that we know that the potential is growing and there is no shift of the null point; however, for any realistic reconstruction, we have to be more careful with such assumptions. 

To conclude, our scheme works well for the non-constant scalar field potential as well---we have checked it with the exponential potential. So, by using the mock data, we have demonstrated that
our scheme works and reconstructs what it should. Now, there are several more points we need to note before testing the scheme with realistic data.

\section{Error Budget and Propagation}
\label{s.err}

\begin{figure}[H]
\includegraphics[width=1.0\textwidth, angle=0]{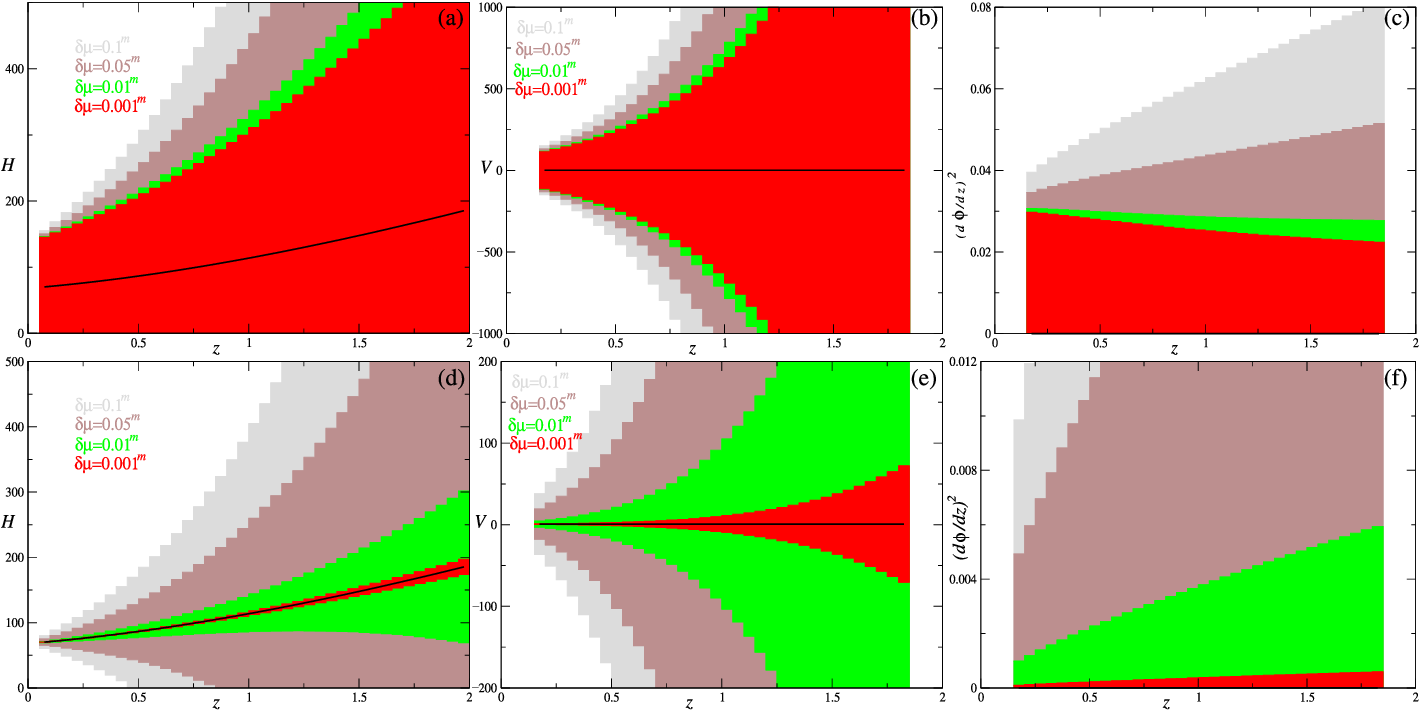}
\caption{{Error} 
budget for $\delta z \ne 0$ ($\delta z = 0.025$ in our particular case) on  panels (\textbf{a}--\textbf{c}) compared to those with $\delta z = 0$ on (\textbf{d}--\textbf{f}) panels. The errors are provided to $H(z)$ on  panels (\textbf{a},\textbf{d}), $V(z)$ on \mbox{ panels} (\textbf{b},\textbf{e}) and
for $(d\phi/dz)^2$ on panels  (\textbf{c},\textbf{f}). Different colors correspond to different $\delta\upmu$: gray to $\delta\upmu=0.1^m$, brown to $\delta\upmu=0.05^m$, green to $\delta\upmu=0.01^m$ and red to $\delta\upmu=0.001^m$
(see the text for more details).}\label{comp}
\end{figure}

As we defined previously, for synthetic data, we use half-bin size as $\delta z$, so that \mbox{$(z_2 - z_1) = 2\delta z$}.  Then, we can substitute it into (\ref{Hz}) and get
\begin{equation}
\begin{array}{l}
\delta (D_M') \cong \dac{\delta D_{M, 2} + \delta D_{M, 1}}{2\delta z} + |D_M'| \dac{2\delta z}{2\delta z} \cong |D_M'| + \mathcal{O}, \\ \\
\delta H = \dac{\delta D_M'}{(D_M')^2} \cong \dac{|D_M'| + \mathcal{O}}{(D_M')^2} = \dac{1}{|D_M'|} + \tilde{\mathcal{O}} = H + \mathcal{O}(H),
\end{array} \label{s.err.1}
\end{equation}
 so that $\delta H \gtrsim H$; similar expressions could be obtained for
$V(z)$ and $(d\phi/dz)^2$. There is some minimum error for the values and this error cannot be lessened  by any means.
The situation is illustrated in Figure~\ref{comp}a--c---one can see that all $H(z)$, $V(z)$ and $(d\phi/dz)^2$ have some ``constant'' contribution to their errors which is not eliminated as
$\delta\upmu\to 0$. We find this situation strange and decided to consider the $\delta z = 0$ case. In this case, the error budget calculations look simpler and the ``constant'' contribution is eliminated; we present the results in
Figure~\ref{comp}d--f. Still, the errors are immense and for the variables of interest ($V(z)$ and $(d\phi/dz)^2$), the errors exceed the values by orders of magnitude. For instance,
$\delta \( (d\phi/dz)^2 \) \sim 5\times 10^{-3}$, as seen from Figure~\ref{comp}f, while $(d\phi/dz)^2 \sim 5\times 10^{-6}$, as seen from Figure~\ref{fig_exp}c, which is 3 orders of magnitude lower;  the situation is similar
with $V(z)$: $V(z) \sim 1$ while $\delta V \sim 100$ (see Figures~\ref{fig_exp}b and \ref{comp}e). This is the situation with synthetic data, the situation with  real data would be even worse.

The main source of increase in the errors lies in the numerical differentiation---indeed, say, $\delta \upmu = 0.01^m$ (green areas in Figure~\ref{comp}) is quite low
(please keep in mind that this is the error of the binned distance modulus)
and unlikely to be achieved, yet,
after \mbox{one differentiation}, it gives rise to a more-or-less wide area in $H(z)$ (see Figure~\ref{comp}d) and after the second differentiation, the resulting error in $V(z)$ is already two orders of magnitude beyond the
central value. So, due to the technique, it is natural to expect big errors in the \mbox{resulting values.}

\begin{figure}[H]
\includegraphics[width=12 cm]{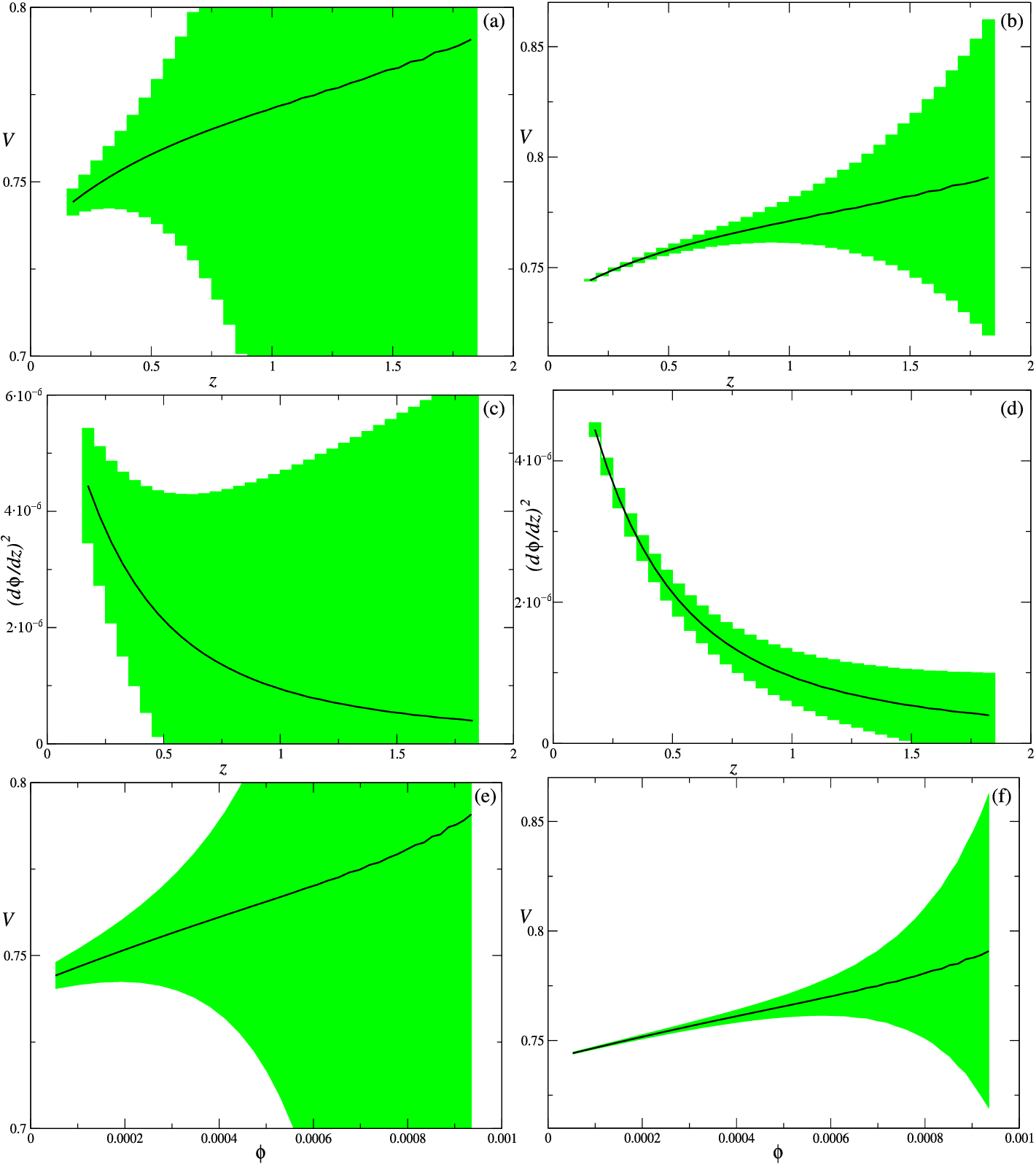}
\caption{{Results} of the reconstruction of potential $V(z)$ (panels (\textbf{a},\textbf{b}) ), kinetic term (~panels (\textbf{c},\textbf{d})) and the potential in the final form $V(\phi)$ ( panels (\textbf{e},\textbf{f})) with use of extremely low values for
$\delta \upmu$: $\delta \upmu = 10^{-5}$ (left column, panels (\textbf{a},\textbf{c},\textbf{e}) ) and $\delta \upmu = 10^{-6}$ (right column, panels (\textbf{b},\textbf{d},\textbf{f}) )
(see the text for more details).}\label{fig_add_2}
\end{figure}

Still, with extremely low values for $\delta \upmu$, a  ``good-looking'' reconstruction of the potential is possible, but the values in question are orders of magnitude below what is reasonable. In Figure \ref{fig_add_2}, we
depict the results of such reconstruction for $\delta \upmu = 10^{-5}$ (left column, panels (a), (c) and (e)) and $\delta \upmu = 10^{-6}$ (right column, panels (b), (d) and (f)). We present a  reconstruction for the
potential $V(z)$ (panels (a) and (b)), kinetic term (panels (c) and (d) ) and the potential in the final form $V(\phi)$ (panels (e) and (f)). One can see that with these low values for $\delta \upmu$, the reconstructed potential
looks quite good. Still, the values for $\delta \upmu$ which are needed for such precision are extremely low and orders of magnitude below \mbox{realistic values.}

\section{Stability of the Reconstruction}
\label{s.stab}

\begin{figure}[H]
\includegraphics[width=11cm]{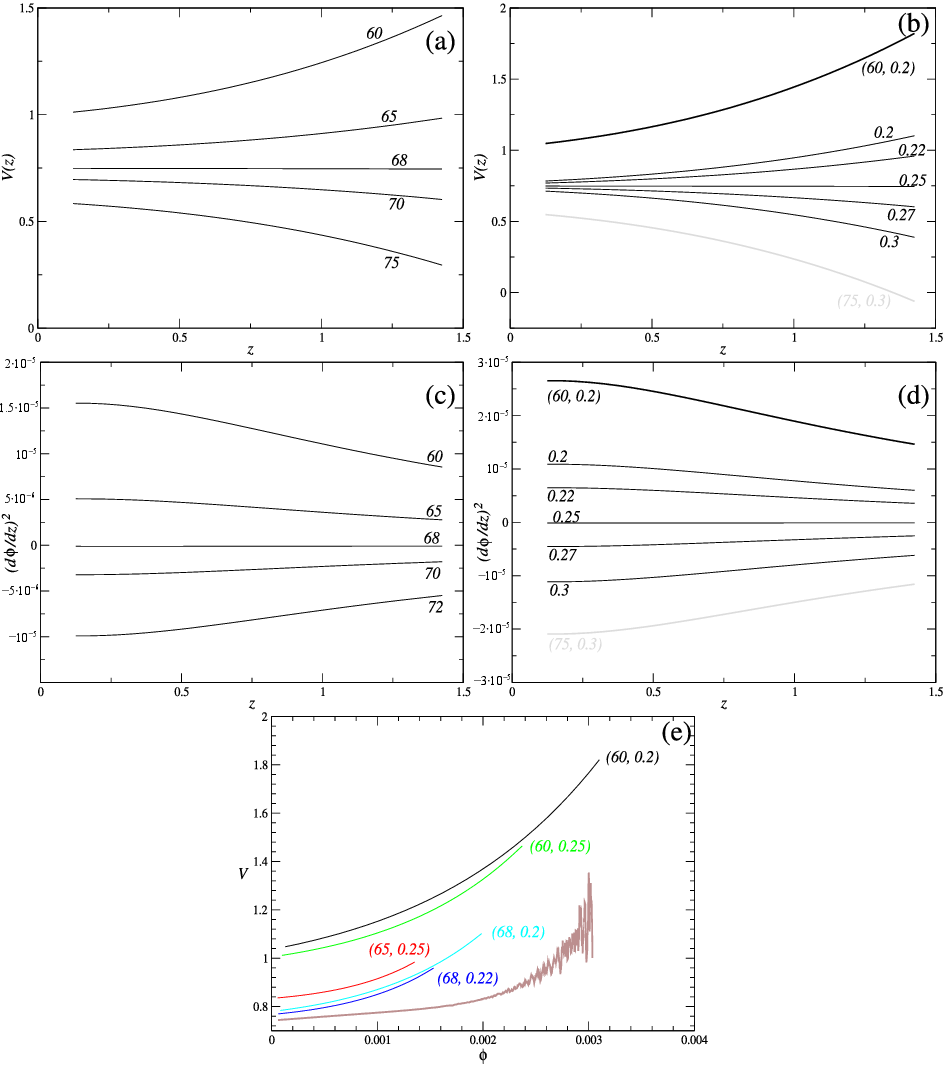}
\caption{{Results} of the potential reconstruction from synthesized $\Lambda$CDM data with $(68, 0.25)$ parameters: potential $V(z)$ reconstruction for different values of $H_0$
(panel (\textbf{a})) and $\Omega_m$ (panel (\textbf{b})); kinetic term reconstruction for different values of $H_0$ (panel (\textbf{c})) and $\Omega_m$ (panel (\textbf{d})); the resulting potential $V(\phi)$ for
viable $(H_0, \Omega_m)$ combinations in  panel (\textbf{e}) and the reconstructed scalar field potential from  Section \ref{sc.f.} as a brown curve
(see the text for more details).}\label{fig6}
\end{figure}

Another important issue we would like to address is the stability of the reconstruction.
As  noted and mentioned in many papers, the $H_0$ and $\Omega_m^0$ are the parameters of the reconstruction and if we use ``wrong'' values for them, the reconstructed
potential will be incorrect. In this section, we investigate the stability of the reconstruction with respect to this. For illustration purposes, we use synthesized $\Lambda$CDM data
with $(H_0, \Omega_m) = (68, 0.25)$ parameters and recover the potential for different values of $H_0$ and $\Omega_m$. The results are presented in Figure~\ref{fig6}. There, in
 panels (a) and (b), we reconstruct the potential $V(z)$ for different values of $H_0$ (in  panel (a)) and $\Omega_m$ (in panel  (b)). One can see that the exact $(68, 0.25)$
values give exactly $\Lambda$CDM with constant potential and a zeroth kinetic term, as demonstrated earlier. One can see that underestimated values for both $H_0$ and $\Omega_m$
give rise to growth of the potential with $z$ while overestimated values start to violate the energy budget. This can be seen more clearly  from panels (c) and (d), where we present
the kinetic term reconstruction. One can see from  panels (c) and (d) that overestimated values for both $H_0$ and $\Omega_m$ result in a negative kinetic term, which is
non-physical (at least within the considered model). Finally, in panel (e), we perform the reconstruction of the potential for underestimated $H_0$ and $\Omega_m$ values. We also present the
reconstructed potential from the scalar field section for comparison (solid brown curve).

So, the  underestimation of the $H_0$ and $\Omega_m$ values acts as effective positive potential while overestimation acts as negative: indeed, we performed the analysis for $\Lambda$CDM
where the ``native'' kinetic term is identically zero, but if there were some ``native'' potential, the overestimation of the $H_0$ and $\Omega_m$ values subtracts some
effective potential from the real one and, depending on the exact potential and the parameters, this could even lead to violation of the energy budget, as in the $\Lambda$CDM case.

The extent of influence of the ``fake'' potential arising from use of the wrong $(H_0, \Omega_m)$ parameters is quite high---as we can see from Figure~\ref{fig6}e, the potentials arising from all the considered
cases exceed the potential reconstructed from our mock scalar field data.

\section{Discussion}
\label{s.discuss}

The main obtained results can be summarized as follows:

\begin{itemize}[leftmargin=7.5mm,labelsep=0.5mm,topsep=3pt]

\item We proposed and developed a scheme which reconstructs the scalar field potential for the Dark Energy without any additional priors or assumptions. The latter is important---indeed, this is not the first
attempt to reconstruct the scalar field potential, but  earlier attempts used priors for $H(z)$ parametrization or the equation of state $w(z)$, or some others; our scheme is free from all of them---we reconstruct the potential purely from raw (synthetized) SNe data (photometric redshift and \mbox{distance modulus);}

\item Our scheme is proved to work perfectly for both $\Lambda$CDM and mock scalar field data;

\item We showed that the resulting uncertainties are huge even for quite precise mock data, and even the limit of 
negligibly small intrinsic uncertainties in data does not eliminate the resulting errors. The reason for this lies in usage of binning---we binned the data and processed the binned data, which implies that if one still wants to reconstruct the potential in an unparametrized way, some other technique should be used;

\item We demonstrated that usage of ``wrong'' (different from apparent) values for $(H_0, \Omega_m)$ gives rise to fake real or ``phantom'' (negative kinetic term) potential. Underestimation of $(H_0, \Omega_m)$
gives rise to real fake potential as shown in Figure~\ref{fig6}e, which could be even greater than the actual potential. Overestimation of $(H_0, \Omega_m)$ gives a negative potential and kinetic term and could
prevent the actual potential from \mbox{being detected.}

\end{itemize}

Let us discuss the results further. The reconstruction method proves to work perfectly on the $\Lambda$CDM data---the recovered $V(z)$ potential is non-zero constant which corresponds to the chosen $\Omega_\Lambda$ and
the kinetic term is zero with  appropriate numerical precision (see Figure~\ref{fig_lcdm}). The exponential potential scalar field data provide for the reconstructed $V(z)$ and the kinetic term, both always positive.
This allows us to further reconstruct the potential in the $V(\varphi)$ form (see Figure~\ref{fig_exp}). One can note a ``flickering'' in the high $z$ region on the reconstructed $V(z)$ and the kinetic term; this comes
from the numerics---the mock data were generated with small $\Delta z$ which leads to``flickering'' of this type while performing numerical differentiation
with even a tiny deviation from the smoothness---and any generated data would be like this. The $H(z)$ curve also has them but of much smaller amplitude, but after performing the second differentiation ($H'(z)$ is
the second derivative of $\upmu(z)$), their amplitude would rise, and that is what we observe in Figures~\ref{fig_exp}b--d. Overall, the resulting reconstructed $V(\phi)$ (see Figure~\ref{fig_exp}d) fits quite well with
the original. As  mentioned in the main text, reconstruction is subject to ambiguity coming from the sign alternating of the square root as well as uncertainty of the reconstructed potential  zero-point location, so that for the reconstruction from  real data, we will need additional insights on both of them. 

In~\cite{pre_proc}, we mentioned that the main source of the extrinsic error is the numerical differentiation---since we are not using any {\emph{ansatz}} 
 for the effective equation of state or $H(z)$ or some other
parametrization, we have to perform full numerical differentiation and this give rise to the uncertainties. To reduce them we decided to skip $\delta z$---errors in $z$. Formally, $z$ is also a result of the measurement,
but different from $\upmu$. Also, while binning, we average data over a number of SNe with different $z$, so consideration of $\delta z \ne 0$ could make sense. But, apparently, consideration of $\delta z \ne 0$ 
equal to the half of the bin size gives rise to a constant contribution to the propagated errors. 
Analytical considerations are given in the appropriate section, and the results are presented in Figure~\ref{comp}.
There, on the upper row ((a--c) panels), we present the results of the errors with $\delta z \ne 0$ for several different $\delta \upmu$. One can easily see that even at $\delta \upmu \to 0$, the resulting error does not
tend to zero but approaches some constant value instead. We found this to be contradictory to common sense and tried the same analysis but with $\delta z = 0$; the results are provided on the bottom row of 
Figures~\ref{comp}d--f.
There, one can see more realistic behavior of the resulting errors for different $\delta \upmu$. Still, the errors for $V(z)$ and the kinetic term are immense and for the considered values for $\delta \upmu$, restoration
of the potential is possible but the resulting uncertainties raise questions about the viability of the reconstruction. And this is happening for $\delta \upmu = 0.1^m \div 0.001^m$---the values are unlikely to be reached
for realistic SNe datasets anytime soon. Still, it is interesting to find values for $\delta \upmu$ which allow clean reconstruction of the potential. So, in Figure~\ref{fig_add_2}, we present reconstruction of the potential for
$\delta \upmu = 10^{-5}$ and $10^{-6}$---one can see the errors for the reconstructed potential, but one also should note that such small values for the $\delta \upmu$ are  impossible from a practical standpoint.

This implies that the binning technique which we used in the current paper to smooth the SNe data amplifies error beyond reasonable values and this method cannot be used for practical purposes. 
Thus, one requires another way around and one of the widely used possibilities is mentioned in the Introduction parametrization of the Hubble parameter. In addition to the {\emph{pro}}s and {\emph{con}}s discussed there, let us add \mbox{one more} \mbox{argument. Generally} speaking, typical parametrization of the Hubble parameter functional form is a mathematical procedure---like that mentioned in the Introduction series on $(1+z)$ powers. However, for the Friedman equations,  this parametrization is made physical with all terms having  physical meaning; so, all the terms of the parametrization should also have some physical meaning and determining this could be nontrivial. 
So, usage of such parametrization, despite being mathematically correct, would raise questions about the physical meaning of the introduced terms. However, if we dig deeper, we will find out that, in some cases, even the usage of parametrization itself is not quite mathematically rigorous---for instance, when using series, they obviously have some radius of convergence and outside of this radius, formally, we cannot consider such parametrization. 

One of the recent potential reconstruction attempts can be found in~\cite{rec_rec}. The authors considered not SNe data but $H(z)$, compiled from different sources such as cosmic chronometers and baryon acoustic oscillations. This simplifies the procedure---only one differentiation is needed in this case, at the cost of a much smaller quantity of data points. Also, to smooth/average the data, the authors used a Gaussian process, and, as a result, the authors reconstructed the potential both in power-law and free forms. 

However, there are examples of physically motivated parametrizations as well---for instance, in~\cite{aadd1}, the authors considered interacting DE and DM with the interaction rate being proportional to the energy density of DE and obtained the functional form of the Hubble parameter as $H^2/H_0^2 = \Omega_{m0} (1+z)^3 + (1-\Omega_{m0})(1+z)^{3(1+w)}$; a similar model but with the interaction rate proportional to the energy density of DM leads to a slightly different \mbox{form \cite{aadd2}} $H^2/H_0^2 = \Omega_{0} (1+z)^3 + (1-\Omega_{0})(1+z)^\alpha$ which allowed~\cite{aadd3} to consider a generalization form of these two. 
However, these two forms are obtained under the condition that DE and DM interact; so, their usage for the ``usual'' model where both DE and DM are conserved individually is not quite rigorous. Overall, one can clearly see that the parametrization indeed allows one to set constraints on the parameters but creates additional questions which are not that easy to address.

Another important issue with the potential reconstruction techniques which we studied is the stability of the reconstruction with respect to the values of $H_0$ and $\Omega_m$. Indeed, as one can see from
(\ref{Vz}) and (\ref{dfdz2}), the values for $H_0$ and $\Omega_m$ are the parameters of the reconstruction, so that they define the shape of the reconstructed potential. So, if we choose them wrongly, we will not be
able to properly reconstruct the potential. To quantitatively analyze the situation, we used $\Lambda$CDM data generated for $H_0 = 68$ km/s/Mpc and $\Omega_m = 0.25$ and reconstructed the potential with different
values for $H_0$ and $\Omega_m$. The results of the reconstruction are presented in Figure~\ref{fig6}. From Figures~\ref{fig6}a,b one can see that underestimation of $H_0$ and $\Omega_m$ leads to the appearance of the
positive potential while underestimation,  the negative. The same is true for the kinetic term, presented in Figures~\ref{fig6}c,d---underestimation of $H_0$ and $\Omega_m$ leads to the appearance of the
positive kinetic term while overestimation to the negative. 
Combining the two, we see that underestimation of $H_0$ and $\Omega_m$ leads to the appearance of the
``fake'' potential while overestimation to its disappearance (if there was any). 
In other words, even if there was no potential to begin with but we underestimate $H_0$ and/or $\Omega_m$, we shall reconstruct a ``fake''
one, while if there is a potential but we overestimate $H_0$ and/or $\Omega_m$, we detect it as suppressed or even do not detect it at all. The magnitude of the ``fake'' reconstructed potential could be estimated from
Figure~\ref{fig6}e where we plot reconstructed potentials from different ($H_0, \Omega_m$) combinations alongside the scalar field potential reconstructed from the mock data (as a brown curve)---one can see that the
amplitude of the ``fake'' potential could easily surpass the real one.
This could be seen as something to be expected---indeed, if we reconstruct with use of parameter values different from ``intrinsic'', one would expect that the result of the reconstruction will not be  the same as the original, but to the best of our knowledge, this is the first quantitative description of the effect so far, making it a good reference.

\section{Conclusions}
\label{s.con}

In this paper, we performed a thorough study of  Dark Energy scalar field reconstruction. We presented a method and tested it with mock data, as well as studied different aspects and the problems which could arise from the
reconstruction process. In our previous paper dedicated to the scalar field reconstruction from SNa Ia data~\cite{pre_proc}, we mentioned that the quality
of the observational data was not enough to enable viable reconstruction; the situation with the observational data is still the same and here we identified possible reasons behind it, including one that is not
so obvious---underestimation of the accepted values for $H_0$ and/or $\Omega_m$. 

The presented method---with no priors/parametrizations for either the Hubble function or the equation of state or other variables---could be seen as a method coming from one side of the methods spectrum. On the opposite side of this spectrum are the methods with such parametrizations. Our method formally allows the reconstruction of any potential of the scalar field while the parametrized approach allows reconstruction only in a specific form. On the other hand, our method generates immense errors while the parametrized reconstructions generate  significantly smaller errors (see the reconstruction in, e.g.,~\cite{rec_rec}). Somewhere between these two approaches lies a method which would optimize the reconstruction at the cost of error budget---and this would be a promising direction for future studies.

We demonstrated that the binning, which we used to smooth the data, introduces errors which far exceeds reasonable values, making the binning technique unacceptable. However, we claim that the unparameterized reconstruction still could be viable. Given that a  smoothing method other than binning may be used, we are currently working in this direction and Gaussian filtering, similar to the approach used in~\cite{add7}, looks promising.

\vspace{6pt}

\supplementary{\textls[-10]{The following supporting information can be downloaded at: \\ \url{https://www.mdpi.com/article/10.3390/universe12070207/s1}}

\noindent Table S1: {\tt synth\_LCDM.dat} (synthetic data for SNe in $\Lambda$CDM model, used in Section~\ref{syndata}); \\
Table S2: {\tt synth\_exp.dat} (synthetic data for SNe in exponential potential model, used in Section~\ref{sc.f.}).
}

\authorcontributions{{ }Data curation, A.P., S.P.; formal analysis, S.P.; investigation, S.P.; methodology, A.P., S.P.; software, A.P., S.P.; visualization, S.P.; writing---original draft, S.P.; writing---review and editing, A.P., S.P.  All authors have read and agreed to the published version of the manuscript.}

\funding{{ } A.P. thanks FAPEMA (project BPV-00040/16) as well as the Higher Education and Science Committee of Armenia (HESCS; grant No. 25IRF/2-1C008) for support. S.P. was supported by FAPEMA under project BPV-00038/16. }

\dataavailability{{ } Data used in this study (synthetic SNe data for $\Lambda$CDM and exponential potential) is available under Supplementary Materials.}

\acknowledgments{{The authors }are deeply grateful to {Luca Amendola} (ITP, Heidelberg U.) for the fruitful discussions, invaluable insights and guidance received throughout the course of this work.
 The authors thank Programa de P\'os-Gradua\c{c}\~ao em F\'isica of the Universidade Federal do Maranh\~ao and {particularly}{ Adalto R. Gomes} for hospitality during their visit where this research was initiated.}

\conflictsofinterest{{ } The authors declare no conflicts of interest. }

\begin{adjustwidth}{-\extralength}{0cm}

\reftitle{References}

\PublishersNote{}
\end{adjustwidth}

\end{document}